\documentclass[12pt,a4paper]{article}
\usepackage{latexsym}
\usepackage{amsfonts}
\usepackage{multicol}
\usepackage[dvips]{graphicx}
\usepackage{fixltx2e}

\newcommand{\danger}[1]{\textbf{#1}}

\input{epsf}             

\begin{document}

\title{\danger{A very simple solution to the OPERA neutrino velocity problem}}
\author{\centerline{\danger{J. Manuel Garc\'\i a-Islas \footnote{
e-mail: jmgislas@leibniz.iimas.unam.mx}}}  \\
Instituto de Investigaciones en Matem\'aticas Aplicadas y en Sistemas \\ 
Universidad Nacional Aut\'onoma de M\'exico, UNAM \\
A. Postal 20-726, 01000, M\'exico DF, M\'exico\\}

\maketitle

\begin{abstract}
Scientists from the OPERA experiment have measured neutrinos supposedly travelling at a 
velocity faster than light contrary to the theory of relativity.
Even when the measurements are precise, the interpretation of
this problem is being misunderstood.
Here it is very easily solved and explained within the theory of
relativity itself proving that neutrinos are not travelling faster than the speed of light 
and the early time arrival is due to the 
the presence of a gravitational field.
\end{abstract}

\section{Introduction}

Scientists from the OPERA experiment \cite{opera} have measured neutrinos supposedly travelling at a 
velocity faster than light therefore in a possibility of contradicting the theory of relativity.
The claim comes from (as cited in their abstract) 
"An early arrival time of CNGS muon neutrinos with respect to the one computed assuming the speed of light in vacuum was measured."

Since the appearance of this claim many papers have given explanations to a possible
solution. Some theoretical 
\cite{cg}, \cite{c}, \cite{h}, and at least one experimental which refutes the superluminal 
velocity \cite{icarus}. Here we give a theoretical solution based completely in the theory
of relativity and
show that this early time arrival is not related to neutrinos travelling faster than the speed
of light but to the time computed from solving the problem in 
the presence of a gravitational field. This simply means that the problem is solved
as taking into account the general theory of relativity.

\section{Particles travelling in the gravitational field.}

The OPERA experiment was situated 1,400 metres below the Earth's surface and neutrinos
were travelling from CERN 731 kilometres away. The neutrinos were travelling inside the Earth
from one city to another.
Two clocks at the emission and reception points were synchronised 
from a satellite.   
Without loss of generality,
the problem of the OPERA experiment can be approximated and stated theoretically in general relativity 
in the following way:

\bigskip

\emph{Suppose that on Earth a massive particle is travelling at velocity $v$ in a circular orbit
(or just in an arc $\Delta \phi$) at a fixed
radial distance $r=R$. Calculate the proper time of travel measured by an 
observer which is fixed at the same radial distance.}

\bigskip

Even when Earth rotates, is not spherical, and the neutrino particles do not travel 
in a circular orbit, the problem can be completely approximated as it has been stated
since for the proof this is enough.
The important thing is the calculation
itself and explain it theoretically as simple as possible.

Let us solve the problem. The general relativity solution around any spherically symmetric 
object(well approximated around the Earth too) is given by the Schwarzschild one.

\begin{equation}
ds^2 = - \bigg( 1- \frac{2M}{r} \bigg) dt^2 + \bigg( 1- \frac{2M}{r} \bigg)^{-1} dr^2 + r^2 (d \theta^2
+ \sin^2 \theta d \phi^2 )
\end{equation}
We use units in which $c =1$ and also $G=1$. A massive particle in Schwarzschild
which travels in a circular orbit $r=R$ at any velocity $v$ has worldline given by

\begin{equation}
x^{\mu} (\tau) = \bigg( \ \gamma \tau \ , \ R \ , \ \frac{\pi}{2} \ , \
\frac{\gamma v}{R} \tau  \ \bigg)
\end{equation}
where 

\begin{equation}
\gamma= \frac{1}{\sqrt{1-v^2 - \frac{2M}{R}}}
\end{equation}
and $\tau$ is a parameter. This parametrisation is such that the 4-velocity vector
satisfies 

\begin{equation}
g_{\mu \nu} \frac{dx^{\mu}}{d\tau} \frac{dx^{\nu}}{d\tau}= -1
\end{equation}

In order for the particle to mantain its motion on the circular
orbit there is need of a radial acceleration. However, we do not need it for our calculation.

From the worldline we have that 

\begin{equation}
\frac{d \phi}{d \tau} = \frac{\gamma v}{R}
\end{equation}

An observer at infinity(satellite) will measure

\begin{equation}
\frac{d \phi}{dt} = \frac{d \phi}{d \tau} \frac{d \tau}{dt}= \frac{v}{R}
\end{equation}
For an observer stationary(fixed at the radial distance $R$) his proper time $dt^{'}$ is related to the
time interval $dt$ by

\begin{equation}
dt^{'} =  \sqrt{\bigg( 1- \frac{2M}{R} \bigg)} \ dt
\end{equation}
and the his proper angle displacement $d\phi^{'}$ is related to $d\phi$ by

\begin{equation}
d\phi^{'} =  R \ d\phi
\end{equation}
Therefore the angular velocity of the particle in orbital motion as measured by the
stationary observer at $R$ is

\begin{equation}
\frac{d \phi^{'}}{dt^{'}} = \frac{R}{ \sqrt{\bigg( 1- \frac{2M}{R} \bigg)}} \ \frac{d \phi}{dt}= 
 \frac{v}{ \sqrt{\bigg( 1- \frac{2M}{R} \bigg)}} 
\end{equation}
The time measured by the stationary observer at $R$ for the particle to travel
an angle displacement $\Delta \phi^{'}$ on Earth is then given by

\begin{equation}
\Delta t^{'} = \frac{\Delta \phi^{'}}{v} \sqrt{\bigg( 1- \frac{2M}{R} \bigg)}
\end{equation}
The experiment was performed below Earth's surface, but the trajectory of each neutrino 
can be approximately thought as a circular arc $\Delta \phi^{'}$, which as a numerical value
is 731 kilometres.  
If the problem of the measured time of the travelling particles is considered only
in a special relativistic way without taking into account the general theory of relativity, 
the time would be $\Delta t^{'} = \Delta \phi^{'} / v$. 
However formula $(10)$ gives the correct
answer of the measured time by a stationary observer when these particles travel an 
angle distance $\Delta \phi^{'}$. 
It can be seen that the time given by formula $(10)$
is in fact shorter than the time the particles will take if they were travelling in flat space.

Therefore the OPERA early arrival time of neutrinos results
are understood not because the particles are travelling
faster than the speed of light(in flat space) but because of formula $(10)$ in curved space.

\end{document}